\documentclass[draft,ef]{agujournal2019}
\usepackage{url} 
\usepackage{lineno}
\usepackage{float}
\usepackage{amsmath}
\usepackage{xcolor}
\usepackage{amsfonts}
\usepackage{gensymb}
\usepackage[inline]{trackchanges} 
\usepackage{soul}
\usepackage{caption}
\usepackage{ragged2e}
\usepackage{subcaption}
\usepackage{makecell}

\draftfalse

\journalname{Earth's Future}

\begin{document}

\title{Deep learning for bias-correcting CMIP6-class Earth system models}

\authors{Philipp Hess\affil{1,2}, Stefan Lange\affil{2}, Christof Schötz\affil{2}, and Niklas Boers\affil{1,2,3}}

\affiliation{1}{\small Earth System Modelling, School of Engineering \& Design, Technical University of Munich, Munich, Germany}
\affiliation{2}{Potsdam Institute for Climate Impact Research, Member of the Leibniz Association, Potsdam, Germany}
\affiliation{3}{Global Systems Institute and Department of Mathematics, University of Exeter, Exeter, UK}

\begin{keypoints}
\item A generative adversarial network is shown to improve daily precipitation fields from a state-of-the-art Earth system model.
\item Biases in long-term temporal distributions are strongly reduced by the generative adversarial network.
\item Our network-based approach can be complemented with quantile mapping to improve precipitation fields further.
\end{keypoints}

%

\justifying

\begin{abstract}
The accurate representation of precipitation in Earth system models (ESMs) is crucial for reliable projections of the ecological and socioeconomic impacts in response to anthropogenic global warming. The complex cross-scale interactions of processes that produce precipitation are challenging to model, however, inducing potentially strong biases in ESM fields, especially regarding extremes. 
State-of-the-art bias correction methods only address errors in the simulated frequency distributions locally at every individual grid cell. Improving unrealistic spatial patterns of the ESM output, which would require spatial context, has not been possible so far.
Here, we show that a post-processing method based on physically constrained generative adversarial networks (cGANs) can correct biases of a state-of-the-art, CMIP6-class ESM both in local frequency distributions and in the spatial patterns at once. While our method improves local frequency distributions equally well as gold-standard bias-adjustment frameworks, it strongly outperforms any existing methods in the correction of spatial patterns, especially in terms of the characteristic spatial intermittency of precipitation extremes. 
\end{abstract}

\section*{Plain Language Summary}
Modelling the impacts of changing precipitation characteristics due to anthropogenic  global warming requires accurate and realistic simulations. These are obtained from Earth system models (ESMs) that solve the governing equations of the atmosphere and oceans on discretized spatial grids, with a resolution of currently around 50-100km. 
Therefore, small-scale physical processes that are crucial for precipitation cannot be simulated directly but have to be included as parameterizations of the resolved variables. This introduces biases that can be adjusted in a post-processing step. 
However, current methods only consider local biases in temporal distributions over long periods and do not account for unrealistic spatial patterns on short-time scales.
Generative adversarial networks are designed to translate unpaired images from one domain to another. Here we use them to generate realistic global high-resolution precipitation fields from ESM simulations and show that the results improve upon those obtained with a state-of-the-art bias correction methodology. 
The results show strongly improved spatial patterns with realistic intermittent characteristics and a similar skill in correcting temporal biases.

\section{Introduction}
Precipitation is a crucial climate variable, and changing amounts, frequencies, or spatial distributions have potentially severe ecological and socioeconomic impacts. With global warming projected to continue in the coming decades, assessing the impacts of changes in precipitation characteristics is an urgent challenge 
\cite{wilcox2007frequency, boyle2010impact, ipcc-ar6-wg1-2021}. Climate impact models are designed to assess the impacts of global warming on, for example, ecosystems, crop yields, vegetation and other land-surface characteristics, infrastructure, water resources, or the economy, in general, \cite{kotz2022effect},
using the output of climate or Earth system models (ESMs) as input. Especially for reliable assessments of the ecological and socioeconomic impacts, accurate ESM precipitation fields to feed the impact models are therefore crucial. 

ESMs are integrated on spatial grids with finite resolution. The resolution is limited by the computational resources that are necessary to perform simulations on decadal to centennial time scales. Current state-of-the-art ESMs have a horizontal resolution on the order of 100km, in exceptional cases going down to 50km. Smaller-scale physical processes that are relevant for the generation of precipitation operate on scales below the size of individual grid cells. These can, therefore, not be resolved explicitly in ESMs and have to be included as parameterizations of the resolved prognostic variables. These include droplet interactions, turbulence, and phase transitions in clouds that play a central role in the generation of precipitation.

The limited grid resolution hence introduces errors in the simulated precipitation fields, leading to biases in short-term spatial patterns and long-term summary statistics. These biases need to be addressed prior to passing the ESM precipitation fields to impact models. In particular, climate impact models are often developed and calibrated with input data from reanalysis data rather than ESM simulations. These reanalyses are created with data assimilation routines and combine various observations with high-resolution weather models. They hence provide a much more realistic input than the ESM simulations and statistical bias correction methods are necessary to remove biases in the ESM simulations output and to make them more similar to the reanalysis data for which the impact models are calibrated. Quantile mapping (QM) is a standard technique to correct systematic errors in ESM simulations. QM estimates a mapping between distributions from historical simulations and observations that can thereafter be applied to future simulations in order to provide more accurate simulated precipitation fields to impact models \cite{deque2007frequency, tong2021bias, gudmundsson2012downscaling, cannon2015bias}.

State-of-the-art bias correction methods such as QM are, however, confined to address errors in the simulated frequency distributions locally, i.e., at every grid cell individually. Unrealistic spatial patterns of the ESM output, which would require spatial context, have so far not been addressed by post-processing methods. For precipitation, this is particularly important because it has characteristic high  intermittency not only in time but also in its spatial patterns.
Multivariate bias correction approaches have recently been developed, aiming to improve spatial dependencies \cite{hess-22-3175-2018, cannon2018multivariate}. However, these approaches are typically only employed in regional studies, as the input dimension becomes too large for global high-resolution ESM simulations. Moreover, such methods have been reported to suffer from instabilities and overfitting, while  differences in their applicability and assumptions make them challenging to use \cite{esd-11-537-2020}.

Artificial neural networks from computer vision and image processing have been successfully applied to various tasks in Earth system science, ranging from weather forecasting \cite{weyn2020improving, rasp2021data, ravuri2021skillful, bi2023accurate, zhang2023skilful},  
to post-processing numerical weather prediction fields \cite{gronquist2021deep, price2022increasing, wang2023customized}, by extracting spatial features with convolutional layers \cite{lecun2015deep}.
Such deep learning-based post-processing methods can be roughly grouped into two categories, using either paired or unpaired training data. \citeA{wang2022deep}, for instance, train a supervised CNN on paired data to correct temperature biases in the output of a ESM from the Climate Model Intercomparison Project phase 6 (CMIP6). On the other hand, \citeA{ravuri2021skillful} use a generative ML model for highly skilful precipitation nowcasting. 

Here, we employ a recently introduced post-processing method \cite{hess2022constrained} that can be trained on unpaired data, based on a cycle-consistent adversarial network (cGAN) to consistently improve both local frequency distributions and spatial patterns of state-of-art CMIP6 ESM precipitation fields.  
Generative adversarial networks \cite{goodfellow2014generative} in particular have emerged as a promising architecture that produces sharp images that are necessary to capture the high-frequency variability of precipitation \cite{ravuri2021skillful, price2022increasing, harris2022generative}. GANs have been specifically developed to be trained on unpaired image datasets \cite{zhu2017unpaired}. This makes them a natural choice for post-processing the output of climate projections, which -- unlike weather forecasts -- are not nudged to follow the trajectory of observations; due to the chaotic nature of the atmosphere, small deviations in the initial conditions or parameters lead to exponentially diverging trajectories \cite{lorenz1996predictability}. As a result, numerical weather forecasts lose their deterministic forecast skill after approximately two weeks at most, and century-scale climate simulations do not agree with observed daily weather records.
Indeed the task of climate models is rather to produce accurate long-term statistics that agree with observations.

We apply our GAN approach to correct global high-resolution precipitation simulations of the GFDL-ESM4 model \cite{gfdlesm4} as a representative CMIP6-class ESM. 
So far, GAN-based approaches have only been applied to post-process ESM simulations either in a regional context \cite{franccois2021adjusting}, or to a very-low-resolution global ESM \cite{hess2022constrained}. 
In the latter study the aim was to improve precipitation fields from an efficient and computationally lightweight ESM, making the combined model a fast ESM with precipitation fields competitive with state-of-the art models. However, the latter study did not show if the cGAN approach could -- in addition to contributing to an efficient, lightweight ESM -- also be used for bias correcting state-of-the-art, CMIP6 class models for subsequent impact modelling. 

Here we show that, indeed,  a suitably designed and trained cGAN can improve even the distributions and spatial patterns of precipitation fields from a state-of-the-art comprehensive ESM, namely GFDL-ESM4. In particular, in contrast to rather specific existing methods for post-processing ESM output for climate impact modelling, we will show that the GAN approach is general and can readily be applied to different ESMs and observational datasets used as ground truth.

In order to ensure that the cGAN-based post-processing does not violate physical conservation laws, we include a suitable physical constraint, enforcing that the cGAN-based transformations do not change the overall global sum of daily precipitation values; essentially, this ensures that precipitation is only spatially redistributed (see Methods). By framing bias correction as an image-to-image translation task, our approach corrects both spatial patterns of daily precipitation fields on short time scales and temporal distributions aggregated over decadal time scales.
We evaluate the skill in improving spatial patterns and temporal distributions against the gold-standard Inter-Sectoral Impact Model Intercomparison Project (ISIMIP) bias adjustment and statistical downscaling method (ISIMIP3BASD) framework \cite{lange2019trend}, which relies strongly on QM. 

Quantifying the ``realisticness'' of spatial precipitation patterns is a key problem in current research \cite{ravuri2021skillful}. We use spatial spectral densities and the fractal dimension of spatial patterns as a measure to quantify the similarity of intermittent and unpaired precipitation fields. 
We will show that our GAN can learn to recognize spatial patterns using convolutional layers and strongly improves the characteristic intermittency in spatial precipitation patterns. We will also show that our GAN, combined with a subsequent application of the ISIMIP3BASD routine, leads to the best overall performance in bias correcting comprehensive ESM fields.
     
\section{Methods}
    \subsection{Training data}
    
       We use global fields of daily precipitation with a horizontal resolution of $1\degree$ from the GFDL-ESM4 Earth system model \cite{gfdlesm4} historical simulation and the W5E5v2 reanalysis product \cite{w5e5v2} as observation-based ground truth. The W5E5v2 dataset is based on the ERA5 \cite{hersbach2020era5} reanalysis and has been bias-adjusted using the Global Precipitation Climatology Centre (GPCC) full data monthly product v2020 \cite{schneider2011gpcc} over land and the Global Precipitation Climatology Project (GPCP) v2.3 dataset \cite{huffman1997global} over the ocean
       to further improve the precipitation statistics of ERA5.
        Both datasets have been regridded to the same $1\degree$ horizontal resolution using bilinear interpolation following \cite{beck2019mswep}. We split the dataset into three periods for training (1950-2000), validation (2001-2003),  and testing (2004-2014). This corresponds to 8030 samples for training, 1095 for validation, and 4015 for testing.
        During pre-processing, the training data is log-transformed with $\tilde{x} = \log(x + \epsilon) - \log(\epsilon)$ with $\epsilon = 0.0001$, following \citeA{rasp2021data}, to account for zeros in the transform. The data is then normalized to the interval $[-1, 1]$ following \cite{zhu2017unpaired}.
    
    \subsection{Cycle-consistent generative adversarial networks}
    
        This section gives a brief overview of the GAN used in this study.
        We refer to \cite{zhu2017unpaired, hess2022constrained} for a more comprehensive description and discussion. 
        Generative adversarial networks learn to generate images that are nearly indistinguishable from real-world examples through a two-player game \cite{goodfellow2014generative}.
        In this setup, a first network $G$, the so-called generator, produces images with the objective of fooling a second network $D$, the discriminator, which has to classify whether a given sample is generated (``fake'') or drawn from a real-world dataset (``real''). Mathematically this can be formalized as 
        
        \begin{equation}
            G^* = \underset{G}{\mathrm{min}} \; \underset{D}{\mathrm{max}} \; \mathcal{L}_{GAN}(D,G),
        \end{equation}
        with $G^*$ being the optimal generator network. The loss function $\mathcal{L}_{GAN}(D,G)$ can be defined as
        
        \begin{equation}
            \mathcal{L}_\mathrm{GAN}(D,G) = \mathbb{E}_{y\sim p_y(y)}[\log ( D(y) )] + \mathbb{E}_{x\sim p_{x}(x)}[\log ( 1- D(G(x)))],
            \label{eq:gan}
        \end{equation}
        where $p_y(y)$ is the distribution of the real-world target data. Samples from the ESM simulation data distribution, here denoted as $p_x(x)$, are used as inputs by $G$ to produce realistic images. 
        The cGAN \cite{zhu2017unpaired} consists of two generator-discriminator pairs, where the generators $G$ and $F$ learn inverse mappings between two domains $X$ and $Y$. 
        This allows defining an additional cycle-consistency loss that constraints the training of the networks, i.e.
        
         \begin{align}
           \mathcal{L}_{\mathrm{cycle}}(G, F) & = \mathbb{E}_{x \sim p_{x}(x)}[|| F(G(x)) - x||_1] \\
           & + \mathbb{E}_{y\sim p_{y}(y)}[|| G(F(y)) - y||_1]. \nonumber
        \end{align}
        
        It measures the error caused by a translation cycle of an image to the other domain and back.
       Further, an additional loss term is introduced to regularize the networks to be close to an identity mapping with
       
         \begin{align}
           \mathcal{L}_{\mathrm{ident}}(G, F) & = \mathbb{E}_{y \sim p_{y}(y)}[|| G(y) - y||_1] \\
           & + \mathbb{E}_{x\sim p_{x}(x)}[|| F(x) - x||_1]. \nonumber
        \end{align}
        The regularization forces the generator networks to output a similar image as given as input to the network. This is in line with the idea that the generator should only improve small-scale features of the image, i.e. change the style, while preserving the large-scale features, i.e. the overall content of the image.
        In practice, the log-likelihood loss can be replaced by a mean squared error loss to facilitate more stable training. Further, the generator loss is reformulated to be minimized by inverting the labels, i.e.
        
        \begin{align}
\mathcal{L}_{\mathrm{Generator}} & =  \mathbb{E}_{x\sim p_{x}(x)}[( D_Y(G(x))-1)^2] \nonumber\\
           & +\mathbb{E}_{y\sim p_{y}(y)}[( D_X(F(y))-1)^2] \\
           & +\lambda \mathcal{L}_{\mathrm{cycle}}(G,F)+\tilde{\lambda} \mathcal{L}_{\mathrm{ident}}(G,F), \nonumber
        \end{align}
        where $\lambda$ and $\tilde{\lambda}$ are set to 10 and 5 respectively following \cite{zhu2017unpaired}.
        The corresponding loss term for the discriminator networks is given by
        
        \begin{align}
\mathcal{L}_{\mathrm{Discriminator}} &  =
            \mathbb{E}_{y\sim p_{y}(y)}[( D_Y(y)-1)^2] + \mathbb{E}_{x\sim p_{x}(x)}[( D_Y(G(x)))^2] \\
           & + \mathbb{E}_{x\sim p_{x}(x)}[( D_X(x)-1)^2] + \mathbb{E}_{y\sim p_{y}(y)}[( D_X(F(y)))^2]. \nonumber
        \end{align}
        The weights of the generator and discriminator networks are then optimized with the ADAM \cite{kingma2014adam} optimizer using a learning rate of $2e^{-4}$ and updated in an alternating fashion.
        We train the network for 350 epochs and a batch size of 1, saving model checkpoints every other epoch. 
        We evaluate the checkpoints on the validation dataset to determine the best model instance.
        
    \subsection{Network Architectures}
    
        Both the generator and discriminator have fully convolutional architectures. The generator uses ReLU activation functions, instance normalization, and reflection padding.
        The discriminator uses leaky ReLU activations with slope 0.2 instead, together with instance normalization. For a more detailed description of the cGAN approach, we refer to our previous study \cite{hess2022constrained}.
        The general network architectures in this study are similar, but with an increase in the number of residual layers in the generator network (the constrained and unconstrained networks) from 6 to 7. 
        
        The final layer of the generator can be constrained to preserve the global sum of the input, i.e. by rescaling
        
        \begin{equation} 
            \tilde{y}_{i} = y_{i} \frac{\sum_i^{N_{\textrm{grid}} } x_{i}}{\sum_i^{N_{\textrm{grid}} }  y_{i}},
            \label{eq:constraint}
        \end{equation}
        where $x_i$ and $y_i$ are grid cell values of the generator input and output, respectively and $N_\textrm{grid}$ is the number of grid cells. The generator without this constraint will be referred to as unconstrained in this study.
        The global physical constraint enforces that the global daily precipitation sum is not affected by the cGAN post-processing and hence remains identical to the original value from the GFDL-ESM4 simulations. This is motivated by the observation that large-scale average trends in precipitation follow the Clausius-Clapeyron relation \cite{traxl2021role}, which is based on thermodynamic relations and hence can be expected to be modelled well in GFDL-ESM4.

    \subsection{Quantile mapping-based bias adjustment}
    
        We compare the performance of our cGAN-based method to the bias adjustment method ISMIP3BASD v3.0.1 \cite{lange2019trend, lange_stefan_2022_6758997} that has
        been developed for phase 3 of the  Inter-Sectoral Impact Model Intercomparison Project (ISIMIP3) \cite{warszawski2014inter, frieler2017assessing}.
        This state-of-the-art bias-adjustment method is based on a trend-preserving quantile mapping (QM) framework.
        It represents a very strong baseline for comparison as it has been developed prior to this study and used not only in ISIMIP3 but also to prepare many of the climate projections that went into the Interactive Atlas produced as part of the 6th assessment report of working group 1 of the Intergovernmental Panel on Climate Change (IPCC, https://interactive-atlas.ipcc.ch/). 

        In the most basic form of QM, cumulative distribution functions (CDFs) of the historical simulation $F_{\mathrm{mod, hist}}(x)$ and observations $F_{\mathrm{obs, hist}}(x)$ are fitted to historical observations and simulations, which can then be applied to bias-adjust future simulations, i.e.,
        
        \begin{equation}
        \hat{x}_{\mathrm{mod, proj}}(t)= F_{\mathrm{obs, hist}}^{-1} [F_{\mathrm{mod, hist}}(x_{\mathrm{mod, proj}}(t))],
        \end{equation}
        
        where $x_{\mathrm{mod, proj}}(t)$ is the model variable at a given grid cell and time instance from a climate projection simulation, and $\hat{x}_{\mathrm{mod, proj}}(t)$ is the quantile-mapped estimate.
        The CDFs used for QM can either be empirical or parametric.

        In the present study we apply parametric QM to the GFDL-ESM4 simulation, and empirical QM to the constrained cGAN output.
        The parametric QM is carried out using ISIMIP3BASD v3.0.1, with settings as specified in \citeA{lange2019trend} for precipitation, i.e., Bernoulli-gamma distributions are used for the CDFs and trends in wet-day precipitation quantiles are preserved multiplicatively in most cases.
        The empirical QM is carried out in the same way as the parametric case, with the difference that empirical instead of parametric distributions are used.
        In both cases (empirical and parametric QM), CDFs are fitted and mapped for each grid cell and day of the year separately, using data from a window of 31 days width centred on the given day of the year for CDF fitting.
        We also tested empirical QM for the GFDL-ESM4 simulation and parametric QM for the constrained cGAN output, but those choices led to slightly worse results.

       To evaluate the methods in this study, we define the grid cell-wise bias as the difference in long-term averages,

       \begin{equation}
           \mathrm{Bias}(\hat{y}, y)  = \frac{1}{T}\sum_{t=1}^T \hat{y}_t - \frac{1}{T}\sum_{t=1}^T y_t,
       \end{equation}

       where $T$ is the number of time steps, $\hat{y}_t$ and  $y_t$ the modelled and observed precipitation respectively at time step $t$.

           \subsection{Evaluating extreme event statistics}
        We are particularly interested in the cGAN's performance in correcting the characteristics of extreme rainfall. To evaluate and benchmark its performance to improve temporal statistics we compare the representation of extreme event statistics in the reanalysis W5E5v2 against the raw GFDL-ESM4 model output and the bias-corrected data, in terms of return values and waiting times.

        \subsubsection{Estimating return values}
        
            The probability of a precipitation event $X$ to exceed a threshold $t \in \mathbb{R}$ is given by $p = 1 - P(X \leq t)$, where $P(X \leq t)$ is the cumulative distribution function (CDF). 
            The return time, i.e., the average number of time steps until an event exceeds the threshold $t$, is then given by $r = 1/p$ and $t$ is called the return value. The aim in the following is to construct the function $t = U(r)$ that maps a return time $r$ to a corresponding return value $t$.

            We employ a peaks-over-threshold method: Denote by $X_{(1)} \leq \dots \leq X_{(n)}$ the local precipitation events ordered by magnitude. Choose $k = \lceil0.05n\rceil$ so that $X_{(n-k+1)}, \dots, X_{(n)}$ exceed the 95th percentile of the empirical distribution of local precipitation events. Following classical results from extreme value theory, e.g., \cite{dehaan06}, we can use these excesses to estimate the tail of their distribution. Here, we apply the so-called moment estimator \cite[Theorem 4.3.1 and equations (3.5.2), (3.5.9), (4.2.4)]{dehaan06},

            \begin{align}
                 M_j &:= \frac1k \sum_{i=0}^{k-1} (\log(X_{(n-i)})-\log(X_{(n-k)}))^j\,,\\
                 \hat{\gamma} &:= M_1 + 1 -\frac12\left(1- \frac{M_1^2}{M_2}\right)^{-1}\,,\\
                 \hat\sigma  &:= \frac12 X_{(n-k)} M_1 \left(1 - \frac{M_1^2}{M_2}\right)^{-1}\,,\\
                 \hat{U}(r) &:= X_{(n-k)} + \hat\sigma \frac{\left( \frac{r k}{n}\right)^{\hat{\gamma}} - 1}{\hat\gamma}\,.\label{eq:return_values}
            \end{align}
            
            The advantage of using the moment estimator over the also common maximum likelihood approach is that the former is more computationally efficient and thus more suitable for the high-resolution fields in our study.
            Using Eq.~\ref{eq:return_values}, we can determine the return value for a given, large return time and a given grid cell to compare the different post-processing methods with regard to their capability of improving these features.
        
        \subsubsection{Waiting times distributions}
        
            Besides return values for a specific return time, i.e. the average waiting time between consecutive events, the \emph{distribution} of waiting times is another key characteristic of extremes and crucial for hydrological impact modelling. We compare the waiting time distributions of the reanalysis and modelled precipitation data by first applying a threshold, defined as the 95th percentile of the local empirical distribution, to the precipitation times series. Counting the number of days between consecutive precipitation events then determines the waiting times. We compare the empirical distributions of the waiting times for each grid cell between the W5E5v2 reanalysis and the different post-processing methods using the Wasserstein-1 distance between distributions, defined as:
            
            \begin{equation}
                W_1(p_x,p_y)  =  \int_{\mathbb{R}} |F_x(z) - F_y(z) | dz,  
            \end{equation}
            
            where $p_x$ and $p_y$ are the distributions of the modelled waiting times and of the reanalysis (W5E5v2) target waiting times respectively, and  $F_x$ and $F_y$, their respective CDFs.

    \subsection{Evaluating spatial patterns}
    
        Quantifying how realistic spatial precipitation fields are, is an ongoing research question in itself, which has become more important with the application of deep learning to weather forecasting and post-processing. 
        In these applications, neural networks often achieve error statistics and skill scores competitive with physical models, while the output fields can, at the same time, show unphysical characteristics, such as blurring or excessive smoothing. \citeA{ravuri2021skillful} compare the spatial intermittency, which is characteristic of precipitation fields, using the radially averaged power spectral density (RAPSD) computed from the spatial fields; in the latter study, the RAPSD-based quantification was complemented by interviews with a large number of meteorological experts. 

        Power spectra provide a useful tool for studying patterns across wide ranges of scales \cite{harris2001multiscale} by applying a Fourier transform that describes the spatial signal in terms of frequencies. 
        The power spectrum of a 2D field $f(x,y)$ can be defined as \cite{ruzanski2011scale}

        \begin{equation*}
            P(f) = |F(k_x,k_y)|^2,
        \end{equation*}
        where
        
        \begin{equation*}
            F(k_x,k_y) = \frac{1}{MN}\sum^{M/2-1}_{x=-M/2}\sum^{N/2-1}_{y=-N/2} f(x,y) e^{-i2\pi \left(\frac{k_x x}{M} + \frac{k_y y}{N} \right)},
        \end{equation*}
        and $M$, $N$ are the dimensions of the field.          
        The radially averaged power spectrum is then found by averaging the power spectrum $P(f)$ in polar coordinates in angular direction as a function of the radius $k = \sqrt{k_x^2 + k_y^2}$, such that

        \begin{equation}
            \bar{P}_r(k) = \frac{1}{N_r(k)} \sum^{N_r(k)}_{i=1} P(k_i),
        \end{equation}
        where $N_r(k)$ is the number of frequency samples and the spatial radial wavelength $\lambda = \Delta/k$ with $\Delta$ is the grid spacing of the field in cartesian coordinates.

\begin{figure}[h!]
          \centering
          \includegraphics[width=1.0\textwidth]{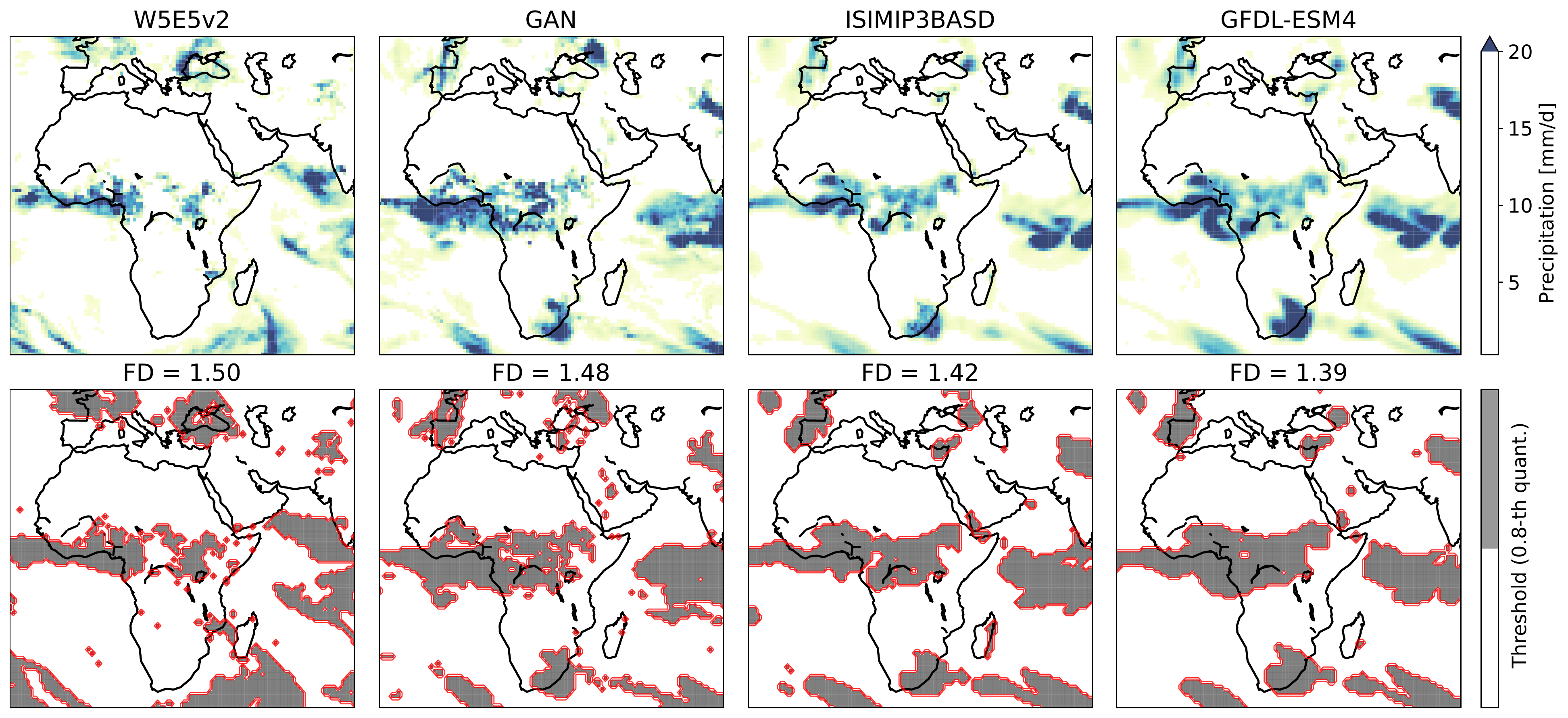}
          \caption{Qualitative comparison of precipitation fields at the same date (31st May 2004). The global fields have been cropped to a region over the African continent to better visualize the small-scale representation of precipitation (see Fig.~S2 and S3 for more examples).  
          The top row shows the daily precipitation fields for the W5E5v2 ground truth, cGAN, ISIMIP3BASD and GFDL-ESM4. The bottom row shows the respective contour lines of binary fields used to compute the fractal dimension (FD) given in the titles.}
          \label{fig:fractal_dimension_local}
        \end{figure}

        We propose here the fractal dimension of precipitation fields as a new metric to quantify how realistic precipitation patterns from generative neural networks are. The fractal dimension has been used in the past already to study how precipitation patterns change across scales \cite{lovejoy1987functional}.
         We compute the fractal dimension via the box-counting algorithm \cite{lovejoy1987functional, meisel1992box, husain2021fractal}.
        The box-counting algorithm divides the image into $N_{\mathrm{sq}}$ squares of linear size $s$ that cover the boundary of binary patterns. We thus first transform the continuous precipitation data into binary fields by applying a global quantile threshold (see Fig.~\ref{fig:fractal_dimension_local} for an example).  
        The size of the squares, i.e. the scale of measurement, is then reduced iteratively by a factor of two and the number of squares or "boxes" are counted at each scale.
        The fractal dimension (FD) can then be determined from the slope of a polynomial fit of the resulting log-log scaling of $N_{\mathrm{sq}}$ as a function of the measured scale with 
        
        \begin{equation}
            \mathrm{FD} = \frac{\log(N_{\mathrm{sq}})}{\log(1/s)}.
        \end{equation}
        
\section{Results}

   We evaluate our cGAN method on two different tasks and time scales. First, the correction of daily rainfall frequency distributions at each grid cell locally, aggregated from decade-long time series. Second, we quantify the ability to improve spatial patterns on daily time scales. Our approach is compared to the raw GFDL-ESM4 model output, as well as to the ISIMIP3BASD methodology applied to the GFDL-ESM4 output.  

    \subsection{Temporal distributions}
        
        \begin{figure}[h]
            \centering
                   \includegraphics[width=0.8\textwidth]{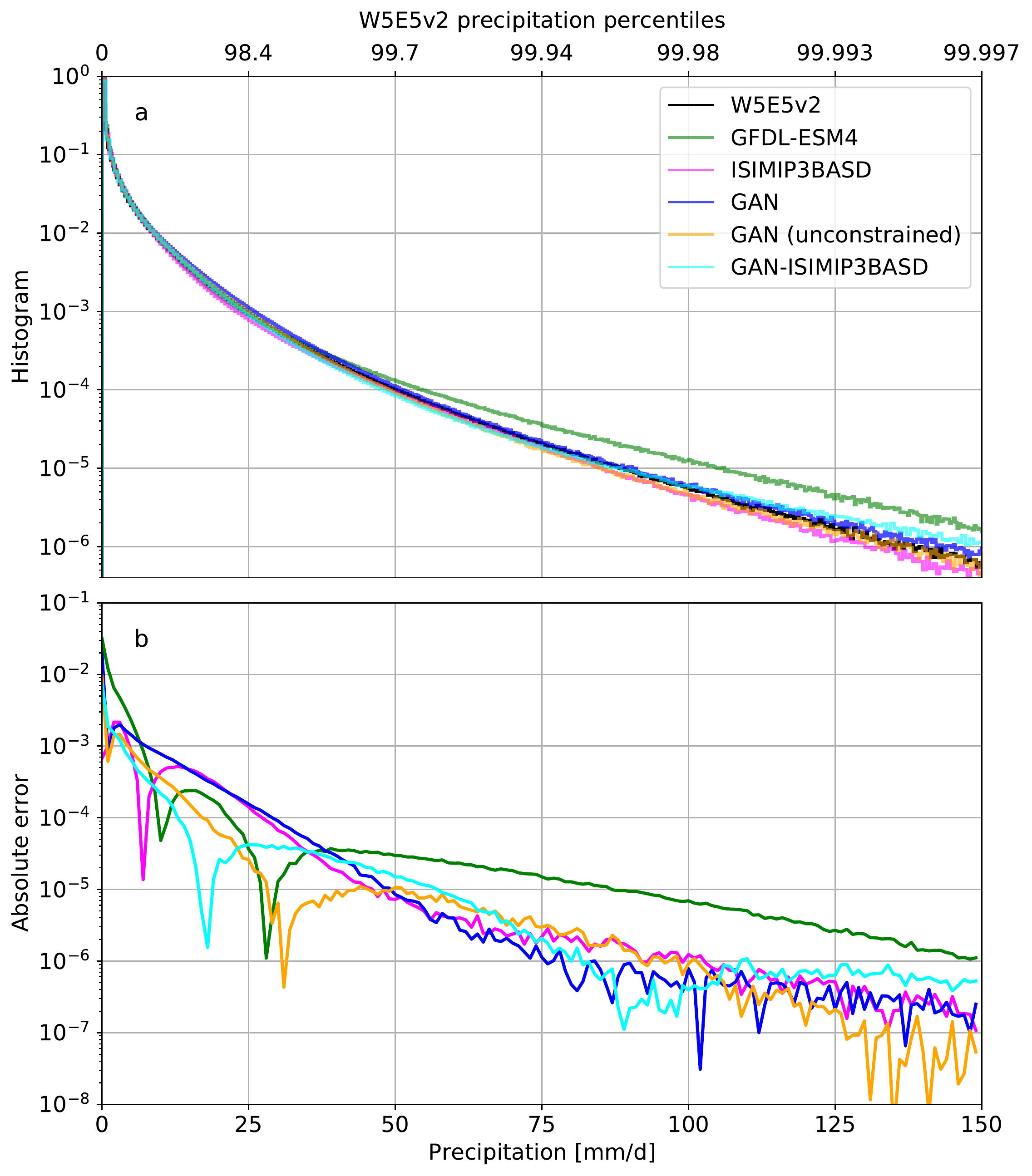}
                \centering
            \caption{Histograms of relative precipitation frequencies over the entire globe and test period (2004-2014). (a) The histograms
            are shown for the W5E5v2 ground truth (black), GFDL-ESM4 (green), ISIMIP3BASD (magenta), cGAN (blue), unconstrained cGAN (orange), and the (constrained) cGAN-ISIMIP3BASD combination (cyan). (b) Distances of the histograms to the W5E5v2 ground truth are shown for the same models as in (a).
            Percentiles corresponding to the W5E5v2 precipitation values are given on the second x-axis at the top. Note that GFDL-ESM4 overestimates the frequencies of strong and extreme rainfall events. All compared methods show similar performance in correcting the local frequency distributions.} 
            \label{fig:histograms}
        \end{figure} 

        We compute global histograms of relative precipitation frequencies using daily time series (Fig.~\ref{fig:histograms}a).
       The GFDL-ESM4 model overestimates frequencies in the tail, namely for events above 50 mm/day (i.e., the 99.7th percentile).
       Our GAN-based method, as well as ISIMIP3BASD and the cGAN-ISIMIP3BASD combination, correct the histogram to match the W5E5v2 ground truth equally well, as can also be seen in the absolute error of the histograms (Fig.~\ref{fig:histograms}b).

\begin{figure}
            \centering
            \includegraphics[width=\textwidth]{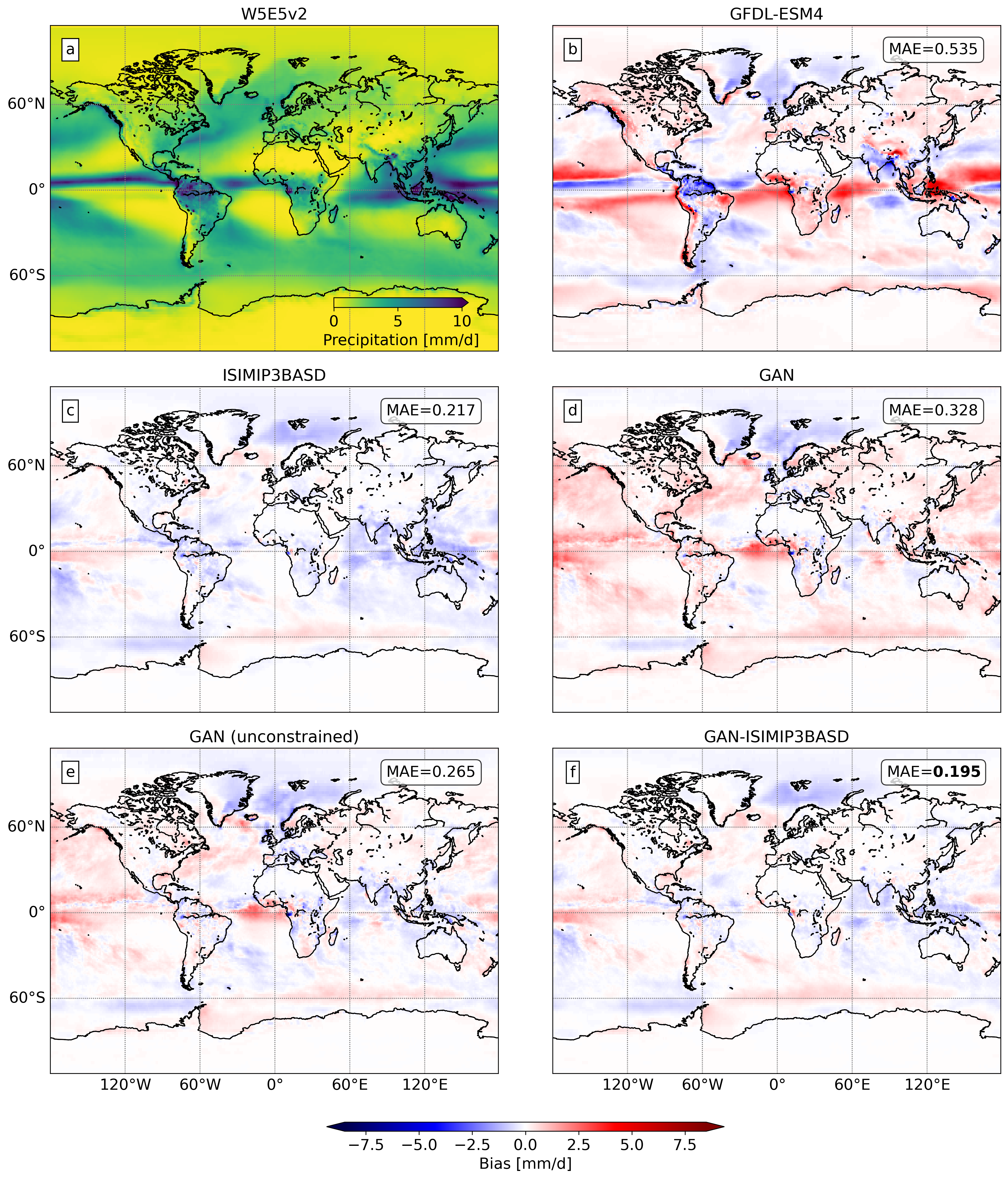}
            \caption{Bias in the long-term average precipitation over the entire test set between the W5E5v2 ground truth (a) and GFDL-ESM4 (b), ISIMIP3BASD (c), cGAN (d), unconstrained cGAN (e) and the (constrained) cGAN-ISIMIP3BASD combination (f). The mean absolute error w.r.t the W5E5v2 ground truth is shown in the upper right corner.}
            \label{fig:bias_maps}
        \end{figure}

\begin{figure}[h!]
            \centering
                   \includegraphics[width=0.9\textwidth]{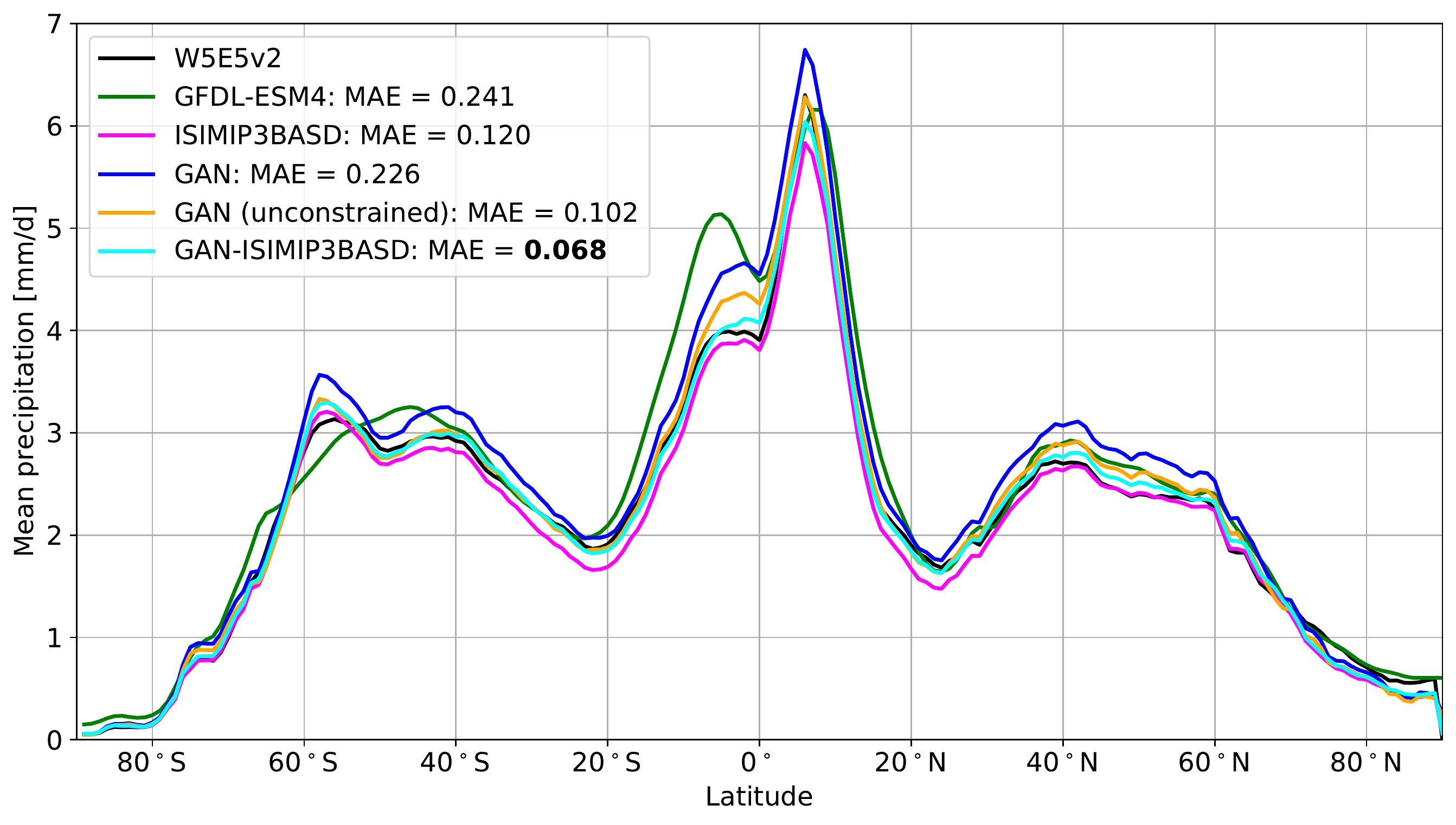}
            \caption{Precipitation averaged over longitudes and the entire test set period from the W5E5v2 ground truth (black) and GFDL-ESM4 (green), ISIMIP3BASD (magenta), cGAN (blue), unconstrained cGAN (orange) and the (constrained) cGAN-ISIMIP3BASD combination (cyan). To quantify the differences between the shown lines, we show their mean absolute error w.r.t the W5E5v2 ground truth in the legend. These values are different from the ones shown in Table \ref{tab:bias} as the average is taken here over the longitudes without their absolute value. The cGAN-ISIMIP3BASD approach shows the lowest error.}
            \label{fig:lat_mean}
        \end{figure}

        Comparing the differences in long-term averages of precipitation per grid cell (Fig.~\ref{fig:bias_maps} and Methods), large biases are apparent in the GFDL-ESM4 model output, especially in the tropics. The double-peaked Inter-Tropical Convergence Zone (ITCZ) bias is visible. The double-ITCZ bias can also be inferred from the latitudinal profile of the precipitation mean in Fig.~\ref{fig:lat_mean}.

        Table~\ref{tab:bias} summarizes the annual biases shown in Fig.~\ref{fig:bias_maps} as absolute averages and additionally for the four seasons. 
        The cGAN alone reduces the annual bias of the GFDL-ESM4 model by $38.7 \%$. The unconstrained cGAN performs better than the physically constrained one, with bias reductions of $50.5\%$. As expected, the ISIMIP3BASD gives even better results for correcting the local mean since it is specifically designed to transform the local frequency distributions accurately. It is, therefore, notable that applying the ISIMIP3BASD procedure on the constrained cGAN output improves the post-processing further, leading to a local bias reduction of the mean by $63.6\%$, compared to ISIMIP3BASD with $59.4\%$. For seasonal time series, the order in which the methods perform is the same as for the annual data.
        
        Besides the error in the mean, we also compute differences in the 95th percentile for each grid cell, shown in Fig.~S1 and as mean absolute errors in Table \ref{tab:bias}. 
        Also, in the case of heavy precipitation values, we find that ISIMIP3BASD improves upon the cGAN but that combining cGAN and ISIMIP3BASD leads to the best agreement of the locally computed quantiles.

         \begin{table}
            \centering
            \footnotesize
            \caption{The globally averaged absolute value of the grid cell-wise difference in the long-term precipitation average, as well as the 95th percentile, between the W5E5v2 ground truth and GFDL-ESM4, ISIMIP3BASD, cGAN, unconstrained cGAN, and the (constrained) cGAN-ISIMIP3BASD combination for annual and seasonal time series (in [mm/day]). The relative improvement over the raw GFDL-ESM4 climate model output is shown as percentages for each method.}
            \begin{tabular}{lccccccccccc}
               \hline
               Season  & \makecell{Percentile} & \makecell{ GFDL-\\ ESM4} & \makecell{ISIMIP3-\\BASD} & \%   & cGAN   & \%    & \makecell{ cGAN \\(unconst.)} & \%    &  \makecell{ cGAN-\\ ISIMIP3-\\BASD} & \% \\ 
               \hline                                                 
               Annual   & -    & 0.535  & 0.217  & 59.4 & 0.328 & 38.7 & 0.265 & 50.5 & \textbf{0.195} & \textbf{63.6} \\
               DJF      & -    & 0.634  & 0.321  & 49.4 & 0.395 & 37.7 & 0.371 & 41.5 & \textbf{0.308} & \textbf{51.4} \\
               MAM      & -    & 0.722  & 0.314  & 56.5 & 0.419 & 42.0 & 0.378 & 47.6 & \textbf{0.285} & \textbf{60.5} \\
               JJA      & -    & 0.743  & 0.289  & 61.1 & 0.451 & 39.3 & 0.357 & 52.0 & \textbf{0.280} & \textbf{62.3} \\
               SON      & -    & 0.643  & 0.327  & 49.1 & 0.409 & 36.4 & 0.362 & 43.7 & \textbf{0.306} & \textbf{52.4} \\
               \hline                                                 
               Annual   & 95th & 2.264  & 1.073  & 52.6 & 1.415 & 37.5 & 1.213 & 46.4 & \textbf{0.945} & \textbf{58.3} \\
               DJF      & 95th & 2.782  & 1.496  & 46.2 & 1.725 & 38.0 & 1.655 & 40.5 & \textbf{1.432} & \textbf{48.5} \\
               MAM      & 95th & 2.948  & 1.482  & 49.7 & 1.805 & 38.8 & 1.661 & 43.7 & \textbf{1.337} & \textbf{54.6} \\
               JJA      & 95th & 2.944  & 1.366  & 53.6 & 1.852 & 37.1 & 1.532 & 48.0 & \textbf{1.247} & \textbf{57.6}\\
               SON      & 95th & 2.689  & 1.495  & 44.4 & 1.741 & 35.3 & 1.592 & 40.8 & \textbf{1.366} & \textbf{49.2} \\
               \hline
            \label{tab:bias}
            \end{tabular}       
        \end{table}

    \subsection{Extremes}

\begin{figure}
            \centering

            \includegraphics[width=\textwidth]{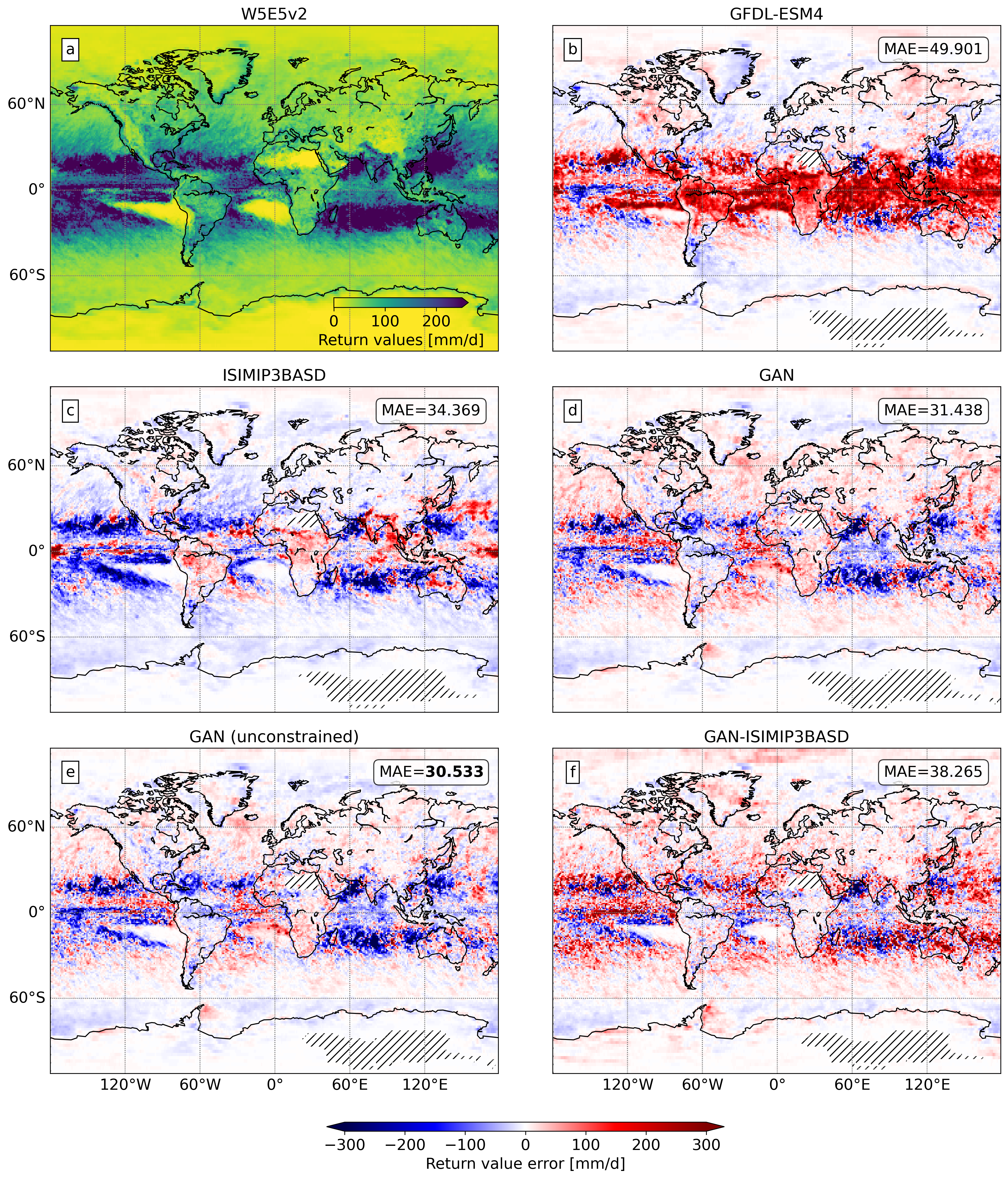}
            \caption{The return values for 10-year return times are shown for the W5E5v2 ground truth (a) and the respective errors for GFDL-ESM4 (b), ISIMIP3BASD (c), the cGAN (d), the unconstrained cGAN (e) and the (constrained) cGAN-ISIMIP3BASD combination (f). 
            The global mean absolute error (MAE) with respect to the W5E5v2 ground truth is shown in the top right corner of each panel, with the unconstrained cGAN exhibiting the lowest error. The hatched area indicates regions that were excluded from the analysis due to insufficient events.}
            \label{fig:return_values}
        \end{figure}

        We evaluate the error of the return values for extreme events with a 10-year return time (see Fig.~\ref{fig:return_values}). The GFDL-ESM4 model shows a strong over-prediction of the return values in the tropics. This is also reflected in the high global mean absolute error (MAE) over the return value differences of 49.9. The ISIMIP3BASD method is able to reduce the MAE to 34.37, but shows a tendency to underestimate the return values. Both the unconstrained and the constrained cGAN alone perform remarkably well, reducing the MAE to 30.53 and 31.44, respectively. The constrained cGAN shows a slight overestimate of return values in the mid-latitudes. 
        Combining the cGAN with the ISIMIP3BASD method shows an overall MAE of 38.27 and does not lead to improved performance (Fig.~\ref{fig:return_values}). 
        We compare latitude profiles of the return value errors, where particularly the cGAN-based post-processings show a significant error reduction in the tropics (Fig.~S5).         
        
\begin{figure}
            \centering
            \includegraphics[width=\textwidth]{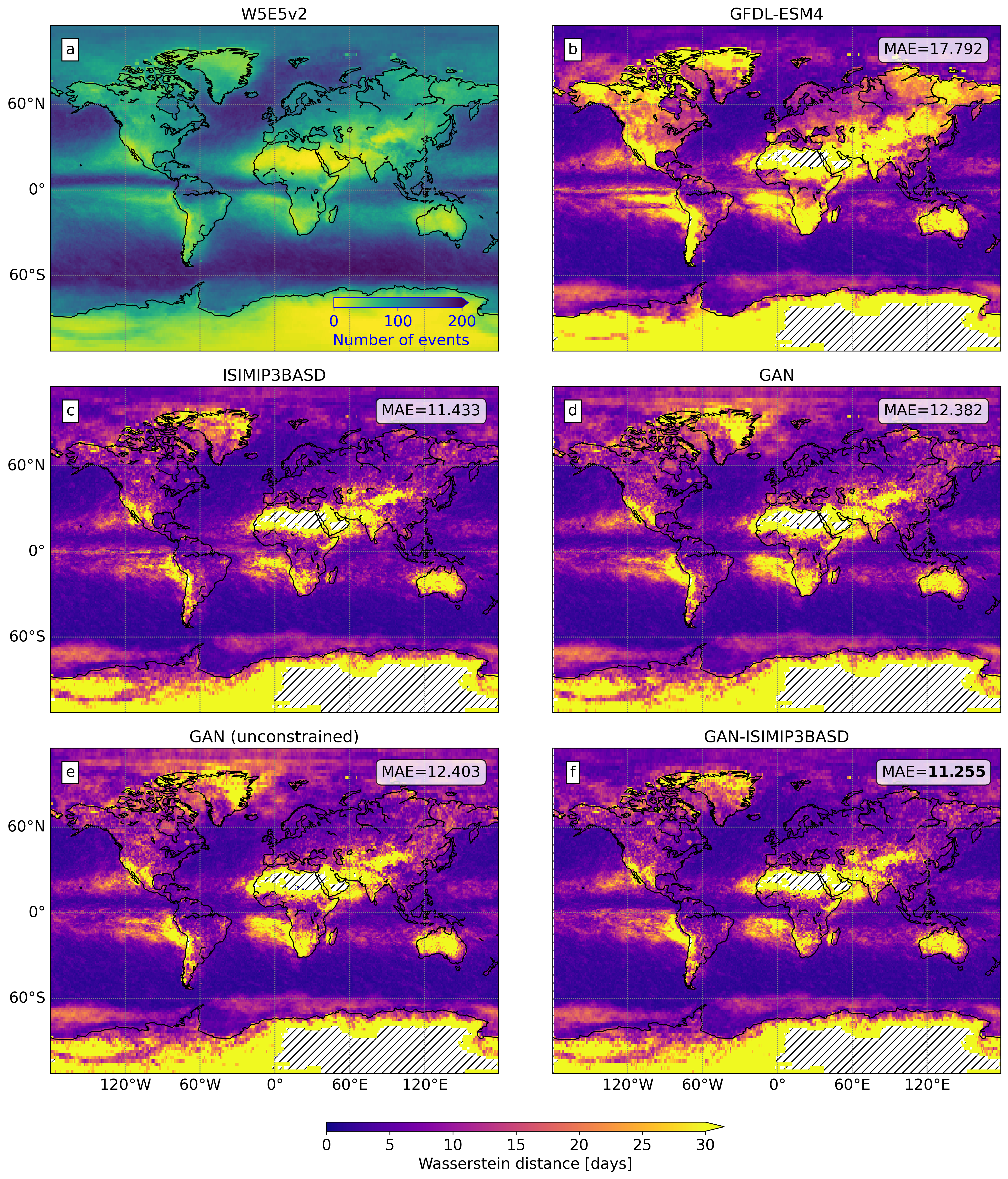}
            \caption{The number of precipitation events above the 95th percentile W5E5v2 ground truth (a) and the relative Wasserstein-1 distance between the waiting time distributions of the ground truth and the GFDL-ESM4 (b), ISIMIP3BASD (c), the constrained cGAN (d), the unconstrained cGAN (e) and the (constrained) cGAN-ISIMIP3BASD combination (f). The global mean absolute error with respect to the W5E5v2 ground truth is shown in the top right corner of each panel, where the cGAN-ISIMIP3BASD method achieves the lowest error. The hatched area indicates regions that were excluded from the analysis due to insufficient events.}
            \label{fig:waiting_times}
        \end{figure}

        To measure the error in the waiting times between extremes, we report the Wasserstein distance between the W5E5v2 reanalysis and the modelled waiting times distributions of precipitation events above the 95th percentile in Fig.~\ref{fig:waiting_times}. Notably, the GFDL-ESM4 model shows a large distance to the ground truth over continental regions, especially in North America and East Asia, which is reduced by all employed post-processing methods. The cGAN-ISIMIP3BASD method improves upon the other approaches with a global MAE of 11.26 in the Wasserstein distance, compared to 17.79 in the case of GFDL-ESM4, 11.43 for ISIMIP3BASD alone, and the constrained and unconstrained cGANs with MAEs of 12.38 and 12.4, respectively. A comparison with the number of events in Fig.~\ref{fig:waiting_times}a shows that the distance in the distributions remains comparably large after correction for regions where the number of precipitation events over the 10-year test period is particularly low. Hence, the error in these regions might either be caused by the models themselves or the insufficient amount of precipitation events. We compare latitude profiles of the Wasserstein distances, which shows a general reduction of distances in the raw GFDL-ESM4 simulation through the post-processing methods, with the best performance given by cGAN-ISIMIP3BASD (Fig.~S6). 
        
    \subsection{Spatial patterns}
    
        We compare the ability of the cGAN to improve spatial patterns based on the W5E5v2 ground truth, against the GFDL-ESM4 simulations and the ISIMIP3BASD method applied to the GFDL-ESM4 simulations. To model realistic precipitation fields, the characteristic spatial intermittency needs to be captured accurately. 
        
\begin{figure}[h!]
          \centering
          \includegraphics[width=1.0\textwidth]{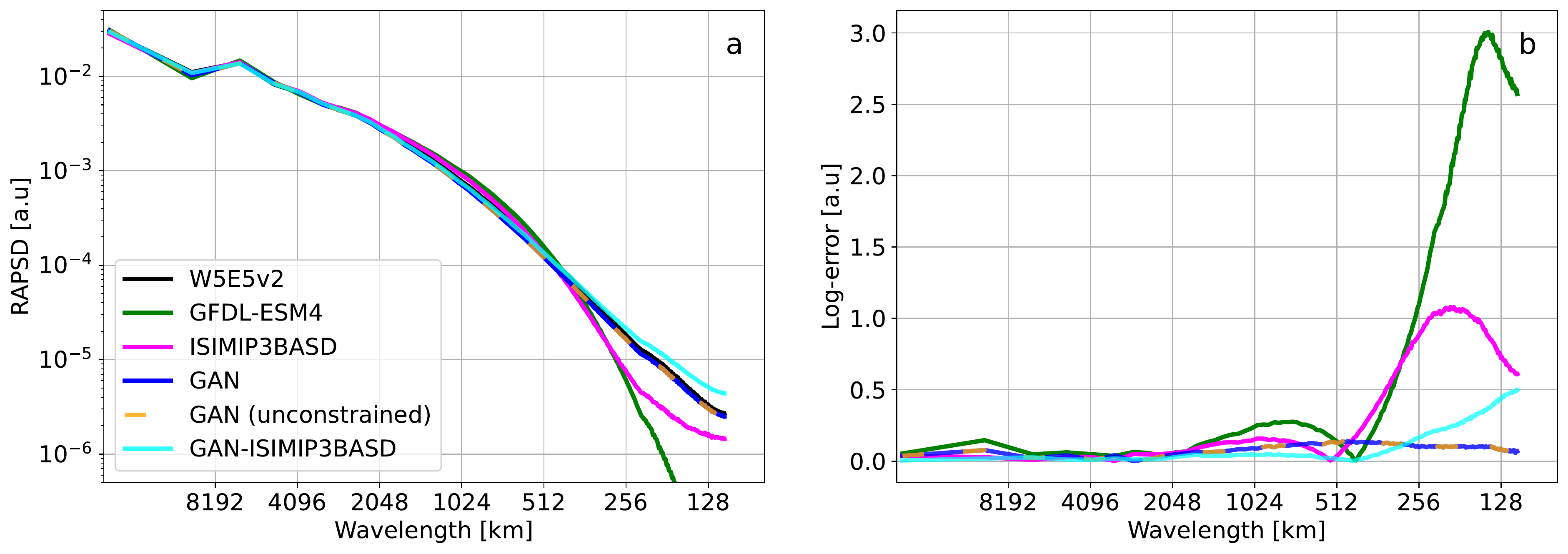}
          \caption{Radially averaged power spectral densities (RAPSDs) of the spatial precipitation fields. (a) The RAPSDs are shown as an average over all samples in the test set, for the W5E5v2 ground truth (black) and GFDL-ESM4 (green), ISIMIP3BASD (magenta), cGAN (blue), unconstrained cGAN (orange, dashed) and the (constrained) cGAN-ISIMIP3BASD
          combination (cyan). 
          (b) The absolute error of the log-transformed spectra w.r.t. the ground truth is shown to highlight the differences.
          The cGANs and W5E5v2 ground truth agree so closely that they are indistinguishable. In contrast to ISIMIP3BASD, the cGAN can correct the intermittent spectrum accurately over the entire range down to the smallest wavelengths.}
          \label{fig:power_spectral_density}
        \end{figure}

        We compute the radially averaged power spectral density (RAPSD) of global precipitation fields, averaged over the test set for each method (Fig.~\ref{fig:power_spectral_density}a). 
        GFDL-ESM4 shows noticeable deviations from W5E5v2 in the RAPSD (Fig.~\ref{fig:power_spectral_density}b). Our cGAN can correct these over the entire range of wavelengths, closely matching the W5E5v2 ground truth. 
        Improvements over ISIMIP3BASD are especially pronounced in the range of high frequencies (low wavelengths), which are responsible for the intermittent spatial variability of daily precipitation fields. 
        Adding the physical constraint to the cGAN does not affect the ability to produce realistic RAPSD distributions. After applying ISIMIP3BASD to the GAN-processed fields, most of the improvements generated by the cGAN are retained, as shown by the cGAN-ISIMIP3BASD results.
        
\begin{figure}[h!]
          \centering
          \includegraphics[width=0.7\textwidth]{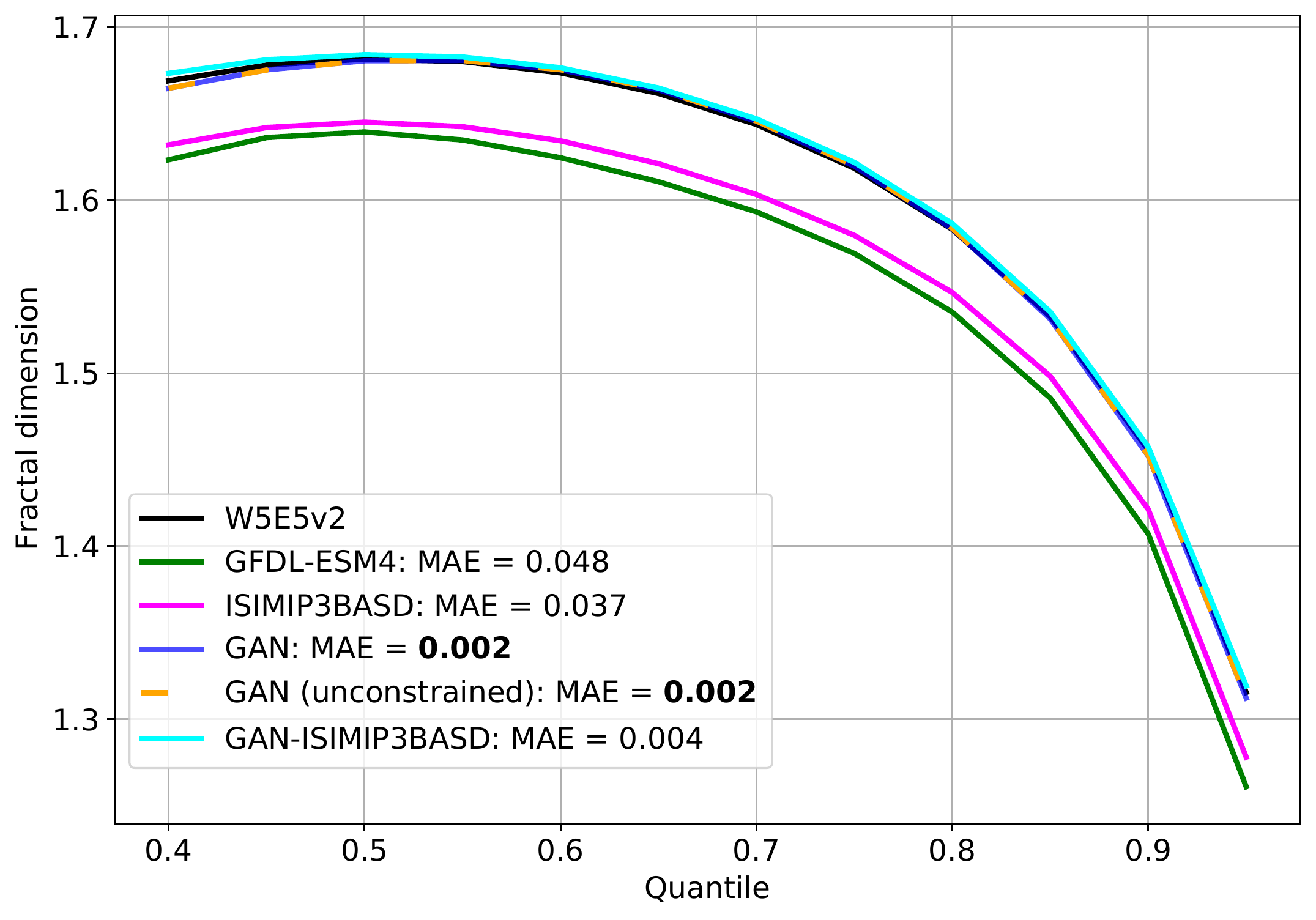}
          \caption{The fractal dimension (see Methods) of binary global precipitation fields is compared as averages for different quantile thresholds. Results are shown for the W5E5v2 ground truth (black) and GFDL-ESM4 (green), ISIMIP3BASD (magenta), cGAN (blue), unconstrained cGAN (orange, dashed), and the (constrained) cGAN-ISIMIP3BASD combination (cyan). The cGAN can accurately reproduce the fractal dimension of the W5E5v2 ground truth spatial precipitation fields over all quantile thresholds, clearly outperforming the ISIMIP3BASD basline.}
          \label{fig:fractal_dimension}
        \end{figure}

        For a second way to quantify how realistic the simulated and post-processed precipitation fields are, with a focus on high-frequency spatial intermittency, we investigate the fractal dimension \cite{edgar2008measure}
        of the lines separating grid cells with daily rainfall sums above and below a given quantile threshold (see Methods). 
        For a sample and qualitative comparison of precipitation fields over the African continent, see Fig.~S2. 
        The daily spatial precipitation fields are first converted to binary images using a quantile threshold.
        The respective quantiles are determined from the precipitation distribution over the entire test set period and globe. 
        The mean of the fractal dimension computed with box-counting (see Methods) \cite{lovejoy1987functional, meisel1992box, husain2021fractal}
        for each time slice is then investigated (Fig.~\ref{fig:fractal_dimension}). 
        Both the GFDL-ESM4 simulations themselves and the results of applying the ISIMIP3BASD post-processing to them exhibit spatial patterns with a lower fractal dimension than the W5E5v2 ground truth, implying too low spatial intermittency. 
        In contrast, the cGAN translates spatial fields simulated by GFDL-ESM4 in a way that results in closely matching fractal dimensions over the entire range of quantiles.

\section{Discussion}
    Post-processing climate projections is a fundamentally different task from post-processing weather forecast simulations \cite{hess2022constrained}. 
    In the latter case, data-driven post-processing methods, e.g. based on deep learning, minimize differences between paired samples of variables such as spatial precipitation fields \cite{hess2022deep}. 
    Beyond time scales of a few days, however, the chaotic nature of the atmosphere leads to exponentially diverging trajectories. For climate or Earth system model output, there is no observation-based ground truth to directly compare to. We, therefore, frame the post-processing of ESM projections with applications for subsequent
    impact modelling in mind as an image-to-image translation task with unpaired samples.
    We focus on precipitation in this study since it is highly relevant for impact assessment in the context of anthropogenic climate change, and because it is arguably the most difficult variable to model due to the high degree of intermittency both in time and in space. However, our method can be extended to multiple variables in a straightforward manner by including them as additional channels in the image.

    To this end, we apply a recently developed post-processing method based on physically constrained GANs to global simulations of a state-of-the-art, high-resolution ESM from the CMIP6 model ensemble, namely the GFDL-ESM4 \cite{gfdlesm4, o2016scenario}. 
    We evaluate our method against the gold-standard bias correction framework ISIMIP3BASD.
     Our model can be trained on unpaired samples that are characteristic of climate simulations. It is able to correct the ESM simulations in two regards: temporal distributions over long time scales, including extremes in the distributions' tails, as well as spatial patterns of individual global snapshots of the model output. The latter is not possible with established methods. Our GAN-based approach is designed as a general framework that can readily apply to different ESMs and observational target datasets. This contrasts existing bias-adjustment methods that are often tailored to specific applications.
        
     We chose to correct precipitation because it is arguably one of the hardest variables to represent accurately in ESMs.  So far, GANs have only been applied to regional studies or low-resolution global ESMs \cite{franccois2021adjusting, hess2022constrained}. 
     The GFDL-ESM4 model simulations are hence chosen in order to test if our cGAN approach would lead to improvements even when post-processing global high-resolution simulations of one of the most complex and sophisticated ESMs to date. In the same spirit, we evaluate our approach against a very strong baseline given by the state-of-the-art bias correction framework ISIMIP3BASD, which is based on a trend-preserving QM method \cite{lange2019trend}.
    
    Comparing long-term summary statistics, our method yields histograms of relative precipitation frequencies that closely agree with corresponding histograms from reanalysis data (Fig.~\ref{fig:histograms}). This means that the extremes in the far end of the tail are accurately captured, with similar skill to the ISIMIP3BASD baseline that is mainly designed for this task. Differences in the grid cell-wise long-term average show that the cGAN skillfully reduces biases (Fig.~\ref{fig:bias_maps}); in particular, the often reported double-peaked ITCZ bias of the GFDL-ESM4 simulations, which is a common feature of most climate models \cite{tian2020double}, is strongly reduced (Fig.~\ref{fig:lat_mean}). The ISIMIP3BASD method - since it is specifically designed for this - produces slightly lower biases for grid-cell-wise averages than the cGAN; we show that combining both methods by first applying the cGAN and then the ISIMIP3BASD procedure leads to the overall best performance. 
    Evaluating the representation of extremes in the bias-adjusted simulations shows that the cGAN outperforms the other models in terms of capturing the magnitude of return values for events with a ten-year return time. 
    We note that applying the ISIMIP3BASD method on cGAN output decreases the performance for 10-year return values. 
    We are able to improve waiting time distributions of precipitation events above the 95th percentile, with the combined cGAN-ISIMIP3BASD method performing better than the other methods.  
    The fact that the combined cGAN-ISIMIP3BASD approach performs better in the latter case could be attributed to the larger set of events available for estimating an accurate mapping of the quantiles. The estimation might be more challenging for the much rarer extremes that only occur once in 10 years on average, given the 20 years of training data.  
    
    Regarding the correction of spatial patterns of the modelled precipitation fields, we compare the spectral density and fractal dimensions of the spatial precipitation fields. Our results show that, indeed, only the cGAN can capture the characteristic spatial intermittency of precipitation closely (Figs.~\ref{fig:power_spectral_density} and \ref{fig:fractal_dimension}). We believe that the measure of fractal dimension is also relevant for other fields, such as nowcasting and medium-range weather forecasting, where blurriness in deep learning-based predictions is often reported \cite{ravuri2021skillful} and needs to be further quantified.
    
    Post-processing methods for climate projections have to be able to preserve the trends that result from the non-stationary dynamics of the Earth system on long-time scales. We have therefore introduced the architecture constraint of preserving the global precipitation amount every day in the climate model output \cite{hess2022constrained}. We find that this does not affect the quality of the spatial patterns that are produced by our GAN method. However, the skill of correcting mean error biases is slightly reduced by the constraint. 
     This can be expected in part as the constraint is constructed to follow the global mean of the ESM (see Fig.~S3). 
     Hence, biases in the global ESM mean can influence the constrained cGAN. 
    This also motivates our choice to demonstrate the combination of the constrained cGAN with the QM-based ISIMIP3BASD procedure since it can be applied to future climate scenarios, making it more suitable for actual applications than the unconstrained architecture. 
    By construction the ISIMIP3BASD method is trend preserving, hence combining it with the constrained cGAN also produces realistic trends (see Fig.~S4).
    In our experiments we also applied QM before the cGAN processing but did not see any notable improvements compared to the cGAN alone.
     
     There are several directions to further develop or approach. The architecture employed here has been built for equally spaced two-dimensional images. Extending the GAN architecture to perform convolutions on the spherical surface, e.g. using graph neural networks, might lead to more efficient and accurate models. 
     Moreover, GANs are comparably difficult to train, which could make it challenging to identify suitable network architectures. 
     Using large ensembles of climate simulations could provide additional training data that could further improve the performance.
     Another straightforward extension of our method would be the inclusion of further input variables or the prediction of additional high-impact physical variables, such as near-surface temperatures that are also important for regional impact models.
        
\section*{Competing interests}
    The authors declare no competing interests.
    
\section*{Open Research}
     The Python code for processing and analysing the data, together with the PyTorch Lightning \cite{pytorch2019} code is available here \cite{github}. 
    The ISIMIP3BASD code in \cite{lange_stefan_2022_6758997} is used for this study.
    The W5E5 data from \cite{w5e5v2} is used and the 
    GFDL-ESM4 data is available here \cite{gfdl_data}.

\acknowledgments

  NB and PH acknowledge funding by the Volkswagen Foundation, as well as the European Regional Development Fund (ERDF), the German Federal Ministry of Education and Research and the Land Brandenburg for supporting this project by providing resources on the high-performance computer system at the Potsdam Institute for Climate Impact Research.
  N.B. acknowledges funding by the European Union's Horizon 2020 research and innovation programme under grant agreement No 820970 and under the Marie Sklodowska-Curie grant agreement No. 956170, as well as from the Federal Ministry of Education and Research under grant No. 01LS2001A.
  SL acknowledges funding from the European Union's Horizon 2022 research and innovation programme under grant agreement no. 101081193 Optimal High Resolution Earth System Models for Exploring Future Climate Changes (OptimESM).
  
\bibliography{literature}

\end{document}